\documentclass[aps,prl,twocolumn,showpacs]{revtex4}
\usepackage{amsfonts,amsmath,amssymb,graphicx,wasysym}
\usepackage{amsfonts}
\usepackage{amsmath}
\usepackage{amssymb}
\usepackage{graphicx}
\usepackage{wasysym}

\def\kbar{\protect\@kbar}
\def\@kbar{\relax \bgroup
\def\@tempa{\hbox{\raise.73\ht0
\hbox to0pt{\kern.25\wd0\vrule width.5\wd0 height.1pt
depth.1pt\hss}\box0}}\mathchoice{\setbox0\hbox{$\displaystyle
k$}\@tempa}{\setbox0\hbox{$\textstyle
k$}\@tempa}{\setbox0\hbox{$\scriptstyle
k$}\@tempa}{\setbox0\hbox{$\scriptscriptstyle k$}\@tempa}\egroup}

\begin{document}

\title{\textbf{Topological Phase Transitions from Harper to Fibonacci Crystals}}
\author{Guy Amit and Itzhack Dana}
\affiliation{Department of Physics, Bar-Ilan University, Ramat-Gan 52900, Israel}

\begin{abstract}
Topological properties of Harper and generalized Fibonacci chains are studied in crystalline cases, i.e., for rational values of the modulation frequency. The Harper and Fibonacci crystals at fixed frequency are connected by an interpolating one-parameter Hamiltonian. As the parameter is varied, one observes topological phase transitions, i.e., changes in the Chern integers of two bands due to the degeneracy of these bands at some parameter value. For small frequency, corresponding to a semiclassical regime, the degeneracies are shown to occur when the average energy of the two bands is approximately equal to the energy of the classical separatrix. Spectral and topological features of the Fibonacci crystal for small frequency leave a clear imprint on the corresponding Hofstadter butterfly for arbitrary frequency.
\end{abstract}

\maketitle
  
\begin{center}
\textbf{I. INTRODUCTION}
\end{center}

Systems having a band spectrum with a nontrivial topological characterization \cite{tknn,ass,bs,ahm,mk,daz,dz,kunz,id0,hk,yh,kz,kz1,kz2,mbb,id1,kz3,kz4,mw,qc1,qc2,qc3,qc4,qc5,qc6,fti,lsz,jg1,id2} are relatively robust to perturbations. Quantum-transport properties determined by the band topological Chern integers, e.g., quantum Hall conductances \cite{tknn,ass,bs,ahm,mk,daz,dz,kunz,id0,hk}, are not altered unless some drastic change occurs in the band spectrum, such as the closing and reopening of a gap as parameters are varied. This is a topological phase transition in which the Chern integers of two bands, which degenerate as the gap closes, change by integer amounts that depend on the system considered and the nature of the degeneracy.

A classic example of a condensed-matter system having a topologically nontrivial band spectrum is that of two-dimensional (2D) crystal electrons in a perpendicular uniform magnetic field. This system was considered in the paper by Thouless \emph{et al.} (TKNN) \cite{tknn}, where the topological characterization of band spectra was introduced. Using this characterization, TKNN explained the quantum Hall effect in a 2D periodic potential for \textquotedblleft rational\textquotedblright\ magnetic fields with flux $\phi =\phi _{0}p/q$ per unit cell, where $\phi _{0}=hc/e$ is the quantum of flux and $(p,q)$ are coprime integers: The contribution of a magnetic band $b$ to the quantum Hall conductance in linear-response theory is $\sigma_{b}e^{2}/h$, where $\sigma_b$ is the Chern integer of the band and satisfies the Diophantine equation \cite{tknn,ahm,daz,dz}: 
\begin{equation}
p\sigma _{b}+q\mu _{b}=1.  \label{de}
\end{equation}
Here $\mu _{b}$ is a second integer. Equation (\ref{de}) holds for general periodic potentials and follows from magnetic (phase-space) translational invariance \cite{daz,dz}. Summing Eq. (\ref{de}) over $N$ filled bands, one gets \cite{daz}:
\begin{equation}
\varphi \sigma +\mu =\rho ,  \label{thc}
\end{equation}
where $\varphi =\phi /\phi _{0}=p/q$, $\rho =N/q$ is the number of electrons per unit cell, and $(\sigma ,\mu )$ are integers with $\sigma e^{2}/h$ being the quantum Hall conductance of the system. Unlike Eq. (\ref{de}), Eq. (\ref{thc}) can be extended to irrational $\varphi $ \cite{daz} by taking the limit of $p$, $q\rightarrow \infty $. For irrational $\varphi$ and for $\rho $ in a gap, Eq. (\ref{thc}) has only one solution $(\sigma ,\mu )$, which is thus universal (system independent) \cite{daz,id1}. In contrast, for rational $\varphi $, Eq. (\ref{thc}) has an infinite number of solutions $(\sigma,\mu)=(\sigma '+rq,\mu '-rp)$, where $(\sigma ',\mu ')$ is some solution and $r$ is any integer. In fact, the value of $\sigma$ (or $\mu$) for rational $\varphi $ is system dependent \cite{tknn,ahm,id0,hk}.

In the case of a periodic potential that is weak relative to the Landau levels spacing, the approximate energy spectrum consists of $p$ magnetic bands splitting from each Landau level \cite{dz}. If the potential is cosinusoidal in one direction, the spectrum is that of a generalized Harper model \cite{hm,ar,dh,aa,djt,hthk,tb0,tb1,tb2}, described by a one-dimensional (1D) tight-binding chain with nearest-neighbor hopping from site $n$ to site $n\pm 1$ and an on-site potential $U(\tau +2\pi\nu n)$; here $U$ is $2\pi$-periodic, $\tau$ is some phase, and the frequency $\nu =1/\varphi$ (see also Sec. II). For irrational $\nu$ (or $\varphi$), this chain is a 1D quasiperiodic system. The nature of the spectrum and the localization properties of the eigenstates of this system depend significantly on $U$ \cite{hm,ar,dh,aa,djt,hthk,tb0,tb1,tb2,tb3,tb4,tb5,tb6,tb7,tb8,tb9,tb10}. Extreme cases are the ordinary Harper model \cite{hm,ar,dh,aa,djt,hthk,tb0}, with a cosine potential $U$, and the generalized Fibonacci quasicrystal \cite{tb0,tb1,tb2,tb3,tb4,tb5,tb6,tb7,tb8,tb9,tb10}, with discontinuous $U$ (see Sec. II). The latter systems are topologically nontrivial \cite{kz,kz1,kz2,mbb,id1,kz3,kz4}, as indicated also by experimental works \cite{kz,kz2,kz3}. However, due to the universality above of the Chern integers $(\sigma ,\mu )$ for irrational $\varphi$ (or $\nu$), the topological properties of these basically different systems are the same, in the sense that open gaps with the same $\rho$ in Eq. (\ref{thc}) for the two systems are labeled by the same values of $(\sigma ,\mu )$ \cite{daz,kz1,id1}. Thus, 1D quasiperiodic systems whose gaps are all open are topologically equivalent. Also, no topological phase transition can occur by the closing and reopening of a gap as parameters are varied. 

For rational $\nu =q/p$, on the other hand, one has a periodic system, a 1D ``crystal", so that the values of $(\sigma ,\mu )$ for given $\rho$ in a gap should depend on $U$ \cite{tknn,ahm,id0,hk}, as mentioned above. Also, in consistency with Eq. (\ref{thc}), $(\sigma ,\mu )$ in periodic systems change generically by $(\pm q,\mp p)$, respectively, at band degeneracies \cite{bs,hk,qc5}. It is then natural to study the topological properties of crystal versions of generalized Harper models, in particular to understand topological phase transitions occurring when $U$ is varied. 

This study is performed in the present paper. The two extreme cases of an ordinary Harper crystal and a generalized Fibonacci crystal are connected by an interpolating Hamiltonian depending on one parameter. Starting from the Harper crystal and gradually increasing the parameter, one approaches the Fibonacci crystal via a sequence of topological phase transitions due to band degeneracies. For small frequency $\nu$, corresponding to a semiclassical regime, the degeneracies are shown to occur when the average energy of the degenerating bands is approximately equal to the energy of the classical separatrix. The band corresponding to the separatrix is in the middle of the spectrum in the case of the Harper crystal, while it is the one just before the highest band in the case of the Fibonacci crystal. In the latter case, the separatrix band is separated from the highest band by the largest spectral gap. These spectral and topological features of the Fibonacci crystal for small frequency are clearly exhibited by the corresponding ``Hofstadter butterfly" (plot of spectra at all frequencies) for arbitrary frequency.

The paper is organized as follows. In Sec. II, we present the model systems to be studied. In Sec. III, we give a background on the basic spectral and topological properties of these systems. In Sec. IV, we consider the semiclassical regime of small frequency. Semiclassical approximations of the band energy spectrum are calculated and compared with the exact spectrum. In Sec. V, we show that the topological properties of the band spectrum reflect the nature of the corresponding classical orbits. We study the topological phase transitions, i.e., the variation of the band Chern integers as the system is gradually changed from an ordinary Harper crystal to a Fibonacci crystal. These transitions are explained using a semiclassical approach. We also study the metamorphosis of the Hofstadter butterfly for ordinary Harper crystals into that for Fibonacci crystals. A summary and conclusions are presented in Sec. VI. 

\begin{center}
\textbf{II. THE MODEL SYSTEMS}
\end{center}   

Consider a 2D crystal potential $V(x,y)$ in a perpendicular uniform magnetic field with $\varphi$ flux quanta per unit cell. At it is well-known \cite{dz,ar,dh} and as detailed in Appendix A for the reader convenience, a sufficiently weak $V(x,y)$ causes a broadening and splitting of a Landau level into an energy spectrum approximately given by that of an effective Hamiltonian $\hat{H}_{\mathrm{eff}}(\hat{X},\hat{Y})$; here the operators $\hat{X}$ and $\hat{Y}$ are the coordinates of the cyclotron orbit center and form a conjugate pair, $[\hat{X},\hat{Y}]=2\pi i\nu$ ($\nu =1/\varphi$). In the case of $V(x,y)=2\lambda W(x)+2\chi\cos(y)$, where $W(x)$ is $2\pi$-periodic and $(\lambda ,\chi)$ are some constants, one finds, by proper choice of $\chi$, that (see Appendix A):
\begin{equation}\label{Heff}
\hat{H}_{\mathrm{eff}}(\hat{X},\hat{Y})=2\lambda U(\hat{X})+2\cos(\hat{Y}),
\end{equation}
where $U(\hat{X})$ is $2\pi$-periodic. As shown in Appendix A, the eigenvalue equation for the operator (\ref{Heff}) in the $\hat{X}$ representation can be written as   
\begin{equation}\label{tbc}
\psi _{n+1}+\psi _{n-1}+2\lambda U(\tau+2\pi n\nu)\psi _{n}=E\psi _{n},
\end{equation}
where $\tau =X_0$ is some arbitrary initial condition. Equation (\ref{tbc}) describes a tight-binding chain with nearest-neighbor hopping and with modulation frequency $\nu$. For irrational (rational) $\nu$, this is a 1D quasiperiodic (periodic) system. Extreme cases of this system are the ordinary Harper model, with $U(X)=\cos (X)$, and the generalized Fibonacci quasicrystal or crystal, with $U(X)=2\left[
\left\lfloor X/(2\pi )+ \nu/2 \right\rfloor -\left\lfloor X/(2\pi )-\nu/2
\right\rfloor \right] -1$, where $\left\lfloor \cdot \right\rfloor $ is the
floor function. The ordinary Fibonacci quasicrystal corresponds to $\nu =(\sqrt{5}-1)/2$. A potential that interpolates between the two cases above is given by a modified version of the one in Ref. \cite{kz1}:
\begin{equation}\label{Ub}
U_{\beta}(X)=\frac{\tanh\{\beta[\cos(X)-\gamma_{\nu}]\}}{\tanh(\beta)}+
\gamma_{\nu}[1-\tanh(\beta)],
\end{equation}
where $\beta$ is a parameter and $\gamma_{\nu}=\cos(\pi\nu)$. It is easy to see that $U_{\beta=0}(X)$ is the ordinary Harper potential while $U_{\beta =\infty}(X)$ is the Fibonacci potential. As shown in Fig. 1, the main features of the Fibonacci potential, i.e., its discontinuities at $X=\pi\nu,\pi (2-\nu )$, are well approximated by formula (\ref{Ub}) for $\beta$ sufficiently large.
\begin{figure}[tbp]
\includegraphics[width=8cm,trim = {0cm 0cm 0cm 0cm}]{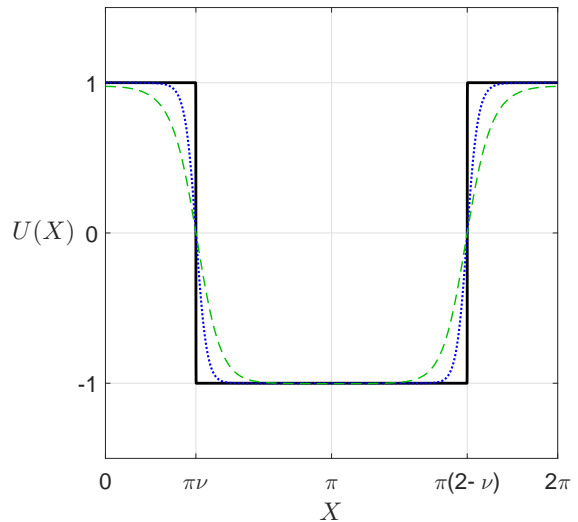}
\caption{(Color online) Fibonacci potential in the basic interval $0\leq X<2\pi$ for $\nu =2/5$ (black solid line) and its approximations by formula (\ref{Ub}) for $\beta =3$ (green dashed line) and for $\beta =7$ (blue dotted line).}
\end{figure}

A well known basic difference between the Harper model and the Fibonacci quasicrystal in the case of $\nu =(\sqrt{5}-1)/2$ is as follows \cite{tb0}: For the Harper model, the states $\psi_n$ in Eq. (\ref{tbc}) at fixed $\tau$ are extended for $\lambda <1$, localized for $\lambda >1$, and critical (poorly localized and not normalizable) for $\lambda =1$; for the Fibonacci quasicrystal, on the other hand, the states are critical for all $\lambda \neq 0$. 

\begin{center}
\textbf{III. BAND EIGENSTATES AND THEIR TOPOLOGICAL PROPERTIES}
\end{center}

\begin{center}
\textbf{IIIA. Band eigenstates and energies}
\end{center}

In the case of rational $\nu$, the system (\ref{tbc}) is periodic and one thus expects a band energy spectrum $E$. As it is well known for similar systems \cite{daz,dz,kz1}, such a spectrum results from the fact that the Hamiltonian (\ref{Heff}) commutes with translations in the $(\hat{X},\hat{Y})$ phase plane and these translations also commute with each other. Then, the simultaneous eigenstates of the Hamiltonian and the phase-plane translations will be Bloch states associated with energy bands. This is shown in some detail in Appendix B and we present here the main results. For rational $\nu =q/p$, a general expression for the Bloch eigenstates of $\hat{H}_{\mathrm{eff}}(\hat{X},\hat{Y})$ in the $X$ representation is
\begin{equation}\label{Bs}
\Psi_{b,{\mathbf k}}(X)=\sum_{m=0}^{p-1}\phi_b(m;{\mathbf k})\Delta_{k_1+2\pi m\nu,k_2}(X).
\end{equation}
Here $b$ is a band index (to be explained below), ${\mathbf k}=(k_1,k_2)$ is a 2D Bloch quasimomentum ranging in the Brillouin zone (BZ)
\begin{equation}\label{BZ}
0\leq k_1 <2\pi\nu,\ \ \ 0\leq k_2 <2\pi /p,
\end{equation}    
$\phi_b(m;{\mathbf k})$ ($m=0,...,p-1$) are $p$ coefficients to be determined, and $\Delta_{\mathbf k}(X)$ are ``$kq$" (Zak) distributions \cite{jz},
\begin{equation}\label{kq}
\Delta_{\mathbf k}(X)=\sum_{j=-\infty}^{\infty}\exp (ijpk_2)\delta (X-k_1-2\pi jq) ,
\end{equation}
forming a complete for $0\leq k_1 <2\pi q$, $0\leq k_2 <2\pi /p$. 
By requiring the state (\ref{Bs}) to be an eigenstate of the effective Hamiltonian (\ref{Heff}) and using the independence of the $p$ $kq$ distributions $\Delta_{k_1+2\pi m\nu,k_2}(X)$ ($m=0,...,p-1$) in Eq. (\ref{Bs}), one can easily derive the eigenvalue equation satisfied by the $p$ coefficients $\phi_b(m;{\mathbf k})$:
\begin{eqnarray}
\phi_b(m-1;{\mathbf k})+\phi_b(m+1;{\mathbf k})&+&2\lambda U(k_1+2\pi m\nu)\phi_b(m;{\mathbf k}) \nonumber \\ &=& E\phi_b(m;{\mathbf k}), \label{eep}
\end{eqnarray}  
$m=0,...,p-1$, with periodic boundary conditions:
\begin{eqnarray}
\phi_b(-1;{\mathbf k})&=&\exp(-ipk_2)\phi_b(p-1;{\mathbf k}), \nonumber \\ \phi_b(p;{\mathbf k})&=&\exp(ipk_2)\phi_b(0;{\mathbf k}). \label{bc}
\end{eqnarray}
While Eqs. (\ref{tbc}) and (\ref{eep}) are similar, no boundary conditions are imposed on Eq. (\ref{tbc}). Equations (\ref{eep}) and (\ref{bc}) define the eigenvalue problem of a ${\mathbf k}$-dependent $p\times p$ matrix $\hat{M}({\mathbf k})$ with column eigenvectors ${\mathbf V}_b({\mathbf k})=\{\phi_b(m;{\mathbf k})\}_{m=0}^{p-1}$. Clearly, this matrix is periodic in  ${\mathbf k}$ in the zone
\begin{equation}\label{BZ1}
0\leq k_1 <2\pi,\ \ \ 0\leq k_2 <2\pi /p.
\end{equation}
For any given value of ${\mathbf k}$ in the zone (\ref{BZ1}), one has $p$ energy eigenvalues $E=E_b({\mathbf k})$, $b=1,...,p$, in Eq. (\ref{eep}). We shall assume that these eigenvalues are all different, i.e., there is no degeneracy. Then, as ${\mathbf k}$ varies in the zone (\ref{BZ1}), these eigenvalues become $p$ isolated (noncrossing) energy bands $E_b({\mathbf k})$. As shown in Appendix B, $E_b({\mathbf k})$ is periodic in both $k_1$ and $k_2$ with period $2\pi /p$, defining a periodicity zone $q$ times smaller than the BZ (\ref{BZ}); this implies that the Bloch states (\ref{Bs}) are $q$-fold degenerate. 

\begin{center}
\textbf{IIIB. Topological Chern integers and Diophantine equation}
\end{center}

For an isolated band $b$, the Bloch eigenstates (\ref{Bs}) must be periodic in the BZ (\ref{BZ}) up to phase factors that may depend on $b$ and on ${\mathbf k}$:
\begin{eqnarray}
\Psi_{b,k_1+2\pi\nu,k_2}(X) &=& \exp [if_b({\mathbf k})]\Psi_{b,{\mathbf k}}(X),\label{pc1} \\
\Psi_{b,k_1,k_2+2\pi /p}(X) &=& \exp [ig_b({\mathbf k})]\Psi_{b,{\mathbf k}}(X),\label{pc2}
\end{eqnarray} 
where $f_b({\mathbf k})$ and $g_b({\mathbf k})$ are the phases. Similarly, the column vector of coefficients ${\mathbf V}_b({\mathbf k})=\{\phi_b(m;{\mathbf k})\}_{m=0}^{p-1}$ in Eq. (\ref{eep}) must be periodic in the zone (\ref{BZ1}) up to phase factors:
\begin{eqnarray}
{\mathbf V}_b(k_1+2\pi ,k_2) &=& \exp [iw_b({\mathbf k})]{\mathbf V}_b({\mathbf k}),\label{vpc1} \\
{\mathbf V}_b(k_1,k_2+2\pi /p) &=& \exp [ig_b({\mathbf k})]{\mathbf V}_b({\mathbf k}),\label{vpc2}
\end{eqnarray}
where $w_b({\mathbf k})$ is another phase and the phase in Eq. (\ref{vpc2}) is the same as that in Eq. (\ref{pc2}) as one can easily verify from Eqs. (\ref{Bs}) and (\ref{kq}). Now, because of the single valuedness of $\Psi_{b,{\mathbf k}}(X)$ in ${\mathbf k}$, the total phase change of $\Psi_{b,{\mathbf k}}(X)$ when going around the boundary of the BZ (\ref{BZ}) counterclockwise must be an integer multiple of $2\pi$. This integer, which we denote by $-\sigma_b$, is a topological characteristic of band $b$ and we shall refer to $\sigma_b$ as a Chern integer. Similarly, the total phase change of ${\mathbf V}_b({\mathbf k})$ when going around the boundary of zone (\ref{BZ1}) (zone BZ1) counterclockwise must be $2\pi\mu_b$, where $\mu_b$ is a second Chern integer. Assuming ${\mathbf V}_b({\mathbf k})$ to be normalized, one can write:
\begin{eqnarray}\label{mub}
& &\mu_b =\frac{1}{2\pi i}\varoint_{\rm BZ1}{\mathbf V}_b^{\dagger}({\mathbf k})\frac{d{\mathbf V}_b({\mathbf k})}{d{\mathbf k}}\cdot d{\mathbf k} \\
&=&\iint_{\rm BZ1}d{\mathbf k}\sum_{b'\neq b}\Im\left\{\frac{{\mathbf V}_b^{\dagger}({\mathbf k})\frac{d\hat{M}^{\dagger}({\mathbf k})}{dk_1}{\mathbf V}_{b'}{\mathbf V}_{b'}^{\dagger}({\mathbf k})\frac{d\hat{M}({\mathbf k})}{dk_2}{\mathbf V}_b}{\pi \left|E_{b'}({\mathbf k})-E_b({\mathbf k})\right|^2} \right\} , \nonumber
\end{eqnarray}           
where $\Im$ denotes imaginary part and $\hat{M}({\mathbf k})$ is the $p\times p$ matrix defined by Eqs. (\ref{eep}) and (\ref{bc}); the sum over $b'\neq b$ in the second line of Eq. (\ref{mub}) (following from the first line by use of Stoke's theorem) is Berry's curvature.

The two Chern integers $\sigma_b$ and $\mu_b$ are connected by the Diophantine equation (\ref{de}). After $\mu_b$ is calculated from Eq. (\ref{mub}), $\sigma_b$ is determined From Eq. (\ref{de}). For the convenience of the reader, Eq. (\ref{de}) is derived in Appendix B.

\begin{center}
\textbf{IV. SEMICLASSICAL APPROXIMATIONS OF BAND SPECTRA}
\end{center}

A good understanding of the topological properties of the band spectrum can be obtained by using semiclassical approximations of this spectrum in order to connect it with orbits of the classical version of the Hamiltonian (\ref{Heff}).  Typical such orbits for the potential (\ref{Ub}) with $\beta =4$ are shown in Figs. 2(a) and 2(b) for two values of $\lambda$. 
\begin{figure}[tbp]
\includegraphics[width=8cm,trim = {1cm 1cm 1cm 1cm}]{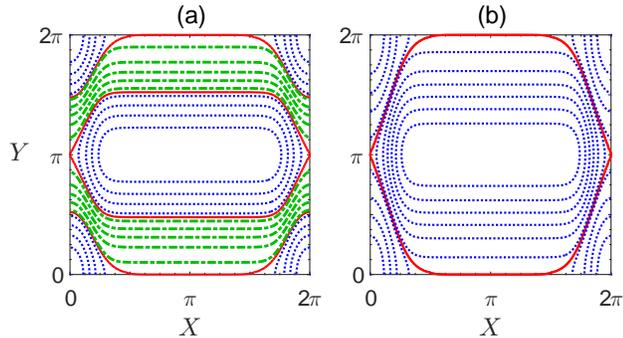}
\caption{(Color online) Classical phase-space diagrams of the Hamiltonian (\ref{Heff}) with potential (\ref{Ub}) for $\nu = 1/11$, $\beta = 4$, and (a) $\lambda$ = 1, (b) $\lambda = \lambda_\beta$, see Eq. (\ref{lc}). Three topologically different kinds of orbits are shown. The blue dotted lines are closed orbits that are contractible to a point. The green dot-dashed lines are open orbits that are not contractible. The red solid lines are the separatrices that are also not contractible. In case (a), one has two separatrices which are separated by open orbits [one separatrix consists of the lowest and uppermost solid lines meeting at $(X=\pi,Y=0)$, equivalent to $(X=\pi,Y=2\pi )$]. In case (b), one has only one separatrix and no open orbits.}
\end{figure}
One can see in Fig. 2(a) three topologically different kinds of orbits: Closed orbits that are contractible to a point, open orbits that are not contractible, and two separatrices or critical orbits, also not contractible. The latter orbits separate between the closed and open orbits. For the special value of $\lambda$ in Fig. 2(b), there are only closed orbits and just one separatrix.  As shown in Appendix C, this case of a single separatrix occurs only for 
\begin{equation}\label{lc}
\lambda=\lambda_{\beta} =\frac{2\tanh(\beta )}{\tanh[\beta(1-\gamma_{\nu})] + \tanh[\beta(1+\gamma_{\nu})]}.
\end{equation}
We also calculate in Appendix C the classical energy of the separatrix for this value of $\lambda$:
\begin{equation}\label{Es}
E_{{\rm S},\beta} =2+2\lambda_{\beta}\left\{\gamma_{\nu}[1-\tanh(\beta)]-\frac{\tanh[\beta(1+\gamma_{\nu})]}{\tanh(\beta)}
\right\}.
\end{equation}
For $\lambda\neq \lambda_{\beta}$ and arbitrary $\beta$, open orbits will be always present. For the sake of definiteness and simplicity, we shall assume from now on the case of Fig. 2(b) ($\lambda=\lambda_{\beta}$).

To connect the band spectrum with the classical orbits, we shall assume a semiclassical regime, i.e., small values of a scaled (dimensionless) Planck constant $\hbar_{\rm s}$. From the relation $[\hat{X},\hat{Y}]=2\pi i\nu$ between the dimensionless conjugate variables $(\hat{X},\hat{Y})$, we see that $\hbar_{\rm s}=2\pi\nu$. Thus, the semiclassical regime is that of small $\nu\ll 1$. In addition, for rational $\nu =q/p$, a classical-quantum correspondence can be established in the simplest way in the ``pure" case \cite{dh} of $q=1$. In fact, the Hamiltonian (\ref{Heff}) is classically invariant under $2\pi$-translations in both $X$ and $Y$ (giving a $2\pi\times 2\pi$ unit cell of periodicity, see Fig. 2). The closest quantum analogue to this invariance is the case when the corresponding quantum translations (\ref{dxy}) commute. From Eq. (\ref{dxdy}), we see that this is possible only for $q=1$, i.e., $\nu =1/p$. We shall assume these values of $\nu$ and also odd $p$ (see note \cite{note}).

For $\nu\ll 1$ or $p\gg 1$, the simplest semiclassical or WKB approximation of energy bands are flat (infinitely degenerate) energy levels $E_l$ ($l$ integer) associated with classical orbits. For the contractible closed orbits in Fig. 2, this association is expressed by the formula
\begin{equation}\label{WKB}
\oint Y(X)dX = 2\pi\hbar_{\rm s}(l+1/2)=4\pi^2\nu (l+1/2),
\end{equation}
i.e., $E_l$ is the energy of the classical orbit whose phase-plane area is given by Eq. (\ref{WKB}) for some integer $l$. The semiclassical energy level associated with the separatrix in Fig. 2(b) is just the classical energy (\ref{Es}). Figure 3 shows, for $\nu =1/p$ ($p=11$) and for several values of $\beta$, the exact energy bands, their semiclassical approximations from Eq. (\ref{WKB}), and the corresponding classical orbits. These results can be understood as follows. Figure 3(a) corresponds essentially to the ordinary Harper model ($\beta =0$) with $\lambda_{\beta}=2$ and $E_{{\rm S},\beta}=0$ from Eqs. (\ref{lc}) and (\ref{Es}). Thus, the separatrix orbit is defined by $\cos (X)+\cos (Y)=0$, i.e., the square of area $2\pi^2$ in Fig. 3(a), which partitions symmetrically the $2\pi\times 2\pi$ unit cell into two regions of equal area. Therefore, the closed orbits satisfying Eq. (\ref{WKB}) can be divided into two groups, each consisting of $(p-1)/2$ orbits. One group is inside the separatrix, surrounding the elliptic fixed point $(\pi ,\pi )$, while the other group is outside the separatrix, surrounding the elliptic fixed point $(0,0)$ and translationally equivalent points. The $p-1$ semiclassical levels associated with these orbits, as well as the exact bands, are then symmetrically positioned below and above the separatrix energy $E=0$.

As $\beta$ is increased, the area of the separatrix region increases, see Fig. 3(b). Then, the number of closed orbits satisfying Eq. (\ref{WKB}) in this region, i.e., the number of semiclassical energy levels below $E_{{\rm S},\beta}$ in Eq. (\ref{Es}), increases beyond $(p-1)/2$. The number of levels above $E_{{\rm S},\beta}$ decreases below $(p-1)/2$.

For $\beta =\infty$, corresponding to the Fibonacci crystal, one has again $\lambda_{\beta}=2$ and $E_{{\rm S},\beta}=0$ from Eqs. (\ref{lc}) and (\ref{Es}). The area of the separatrix region is maximal, see Fig. 3(c). This region and all the orbits inside it lie within the interval $\pi /p< X\leq 2\pi-\pi /p$, where the potential $U(X)=-1$ (see Fig. 1). Thus, from Eq. (\ref{Heff}) with $H_{\mathrm{eff}}$ equal to the orbit energy $E$, we find that each orbit assumes just two $Y$ values in the $X$ interval above:
\begin{equation}\label{Y}
Y_{1,2}=\arccos(E/2+1),\ \ 2\pi -\arccos(E/2+1). 
\end{equation}
Therefore, each orbit is the boundary of the rectangle $\pi /p< X\leq 2\pi-\pi /p$, $Y_1\leq Y\leq Y_2$. One can then explicitly calculate the area in Eq. (\ref{WKB}) for $\nu =1/p$:
\begin{equation}\label{intFC}
\oint Y(X)dX =2\pi(1-1/p) [2\pi -2\arccos(E/2+1)]. 
\end{equation} 
Using Eq. (\ref{intFC}) in Eq. (\ref{WKB}), we obtain the semiclassical energy levels
\begin{equation}\label{El}
E_l=-2-2\cos\left[\frac{\pi}{p-1}\left( l+\frac{1}{2}\right)\right],
\end{equation}
where $l=0,...,p-2$, since for $l>p-2$ the area (\ref{WKB}) is larger than that of the separatrix region, $4\pi^2(1-1/p)$. The level $E_{p-2}\approx 0$ should correspond to the separatrix and will be replaced by the classical energy $E=0$.
Since there can be only $p$ levels approximating the $p$ bands, there remains
only one level to find and this must correspond to an orbit outside the separatrix, in the intervals $0< X\leq \pi /p$ and $2\pi -\pi /p< X\leq 2\pi$, where $U(X)=1$, see Fig. 3(c). Proceeding as above, we now find that $\arccos(E/2-1)=\pi (l+1/2)$, giving the remaining level:
\begin{equation}\label{Ep}
E_{p-1}=2.
\end{equation}
We note that all $p-1$ levels (\ref{El}) lie in the interval $-4<E_l<0$, so that the average gap between neighboring levels, $\Delta E\approx 4/(p-1)$, vanishes as $p\rightarrow \infty$. On the other hand, the highest and largest gap, between the separatrix energy $E=0$ and the last level (\ref{Ep}), is independent of $p$. 
     
\begin{figure}[tbp]
\includegraphics[width=8.2cm,trim = {0cm 1cm 0cm 1.5cm}]{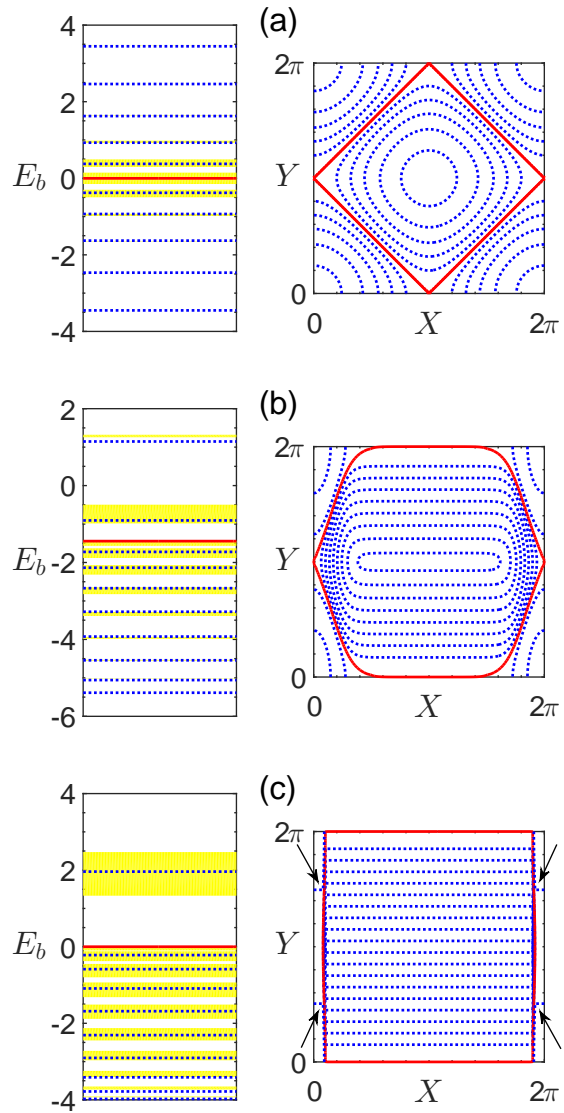}
\caption{(Color online) Energy bands and their semiclassical-level approximations (left diagrams) and classical phase spaces (right diagrams) for $\nu = 1/11$, $\lambda = \lambda_\beta$, and (a) $\beta = 0.01$, (b) $\beta = 4$, and (c) $\beta = 500$. In the left diagrams, the yellow (light gray) regions are the energy bands, the blue dotted lines are the semiclassical levels, and the red solid line is the separatrix energy. In the right diagrams, the red solid line is the separatrix and the blue dotted lines are the classical orbits corresponding to the levels in the left diagram. As $\beta$ is increased, the area of the separatrix region and the number of orbits inside this region increase. In case (c), corresponding essentially to the Fibonacci crystal ($\beta =\infty$), the area of the separatrix region is maximal and there is only one quantized orbit outside this region, indicated by arrows.}
\end{figure}

\begin{center}
\textbf{V. TOPOLOGICAL PHASE TRANSITIONS}
\end{center}

Using Eqs. (\ref{de}) and (\ref{mub}), we have calculated the Chern integers $\sigma_b$ of the bands for rational values of $\nu$ and for $\beta$ varying from $\beta =0$ (ordinary Harper model) to very large values corresponding essentially to the Fibonacci crystal. For $\nu =1/p$ ($q=1$) and for all $\beta$, we found that only one band has a nonzero Chern integer, $\sigma_b=1$; for the other $p-1$ bands, $\sigma_b=0$. Fig. 4 shows results for $p=11$. The solid line is the classical energy (\ref{Es}) of the separatrix versus $\beta$. It can be seen that as $\beta$ is varied the band with $\sigma_b=1$ is always the one associated semiclassically with the separatrix. All the other bands, associated with closed orbits inside or outside the separatrix region (see Fig. 3), have $\sigma_b=0$. This can be understood from the fact that the separatrix orbit is not contractible to a point, since it extends over all a torus, i.e., the $2\pi \times 2\pi$ unit cell of periodicity, in both the $X$ and $Y$ directions. On the other hand, all other closed orbits are not extended, being localized inside the unit cell and therefore contractible to a point. This topological difference between the separatrix and other orbits manifests itself in the nonzero value of $\sigma_b=1$ for the separatrix band, in contrast with $\sigma_b=0$ for the other bands.

\begin{figure}[tbp]
\includegraphics[width=8.2cm,trim = {1cm 1cm 0cm 0cm}]{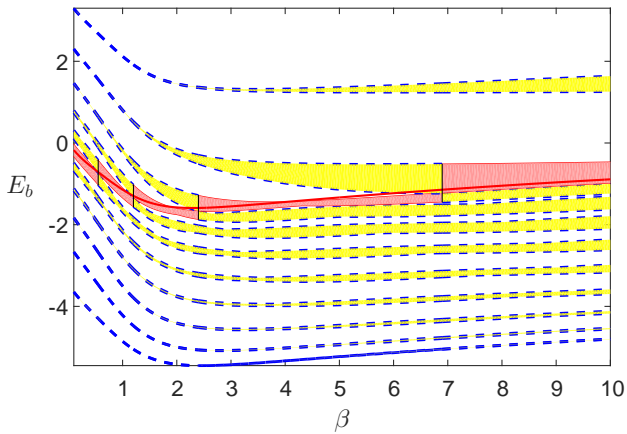}
\caption{(Color online) Energy bands as function of $\beta$ for $\nu = 1/11$ and $\lambda = \lambda_\beta$. The bands corresponding to contractible orbits, with Chern integer $\sigma_b=0$, are the yellow (light gray) regions bounded by blue dashed lines. The bands corresponding to the separatrix, with $\sigma_b=1$, are the red (gray) regions bounded by red solid lines. The thick red solid line is the separatrix energy. Values of $\beta$ where band degeneracies occur are indicated by vertical black segments.}
\end{figure}

As we have shown in Sec. IV (see also Fig. 3), the position of the separatrix band relative to the other bands varies with $\beta$ and is approximately given by Eq. (\ref{Es}). For $\beta =0$, corresponding to the ordinary Harper model, this band is $b=(p+1)/2$, in the middle of the spectrum, see Fig. 3(a). As $\beta$ is increased, the value of $b$ for the separatrix band increases beyond $(p+1)/2$, see Fig. 3(b). For very large $\beta$ (Fibonacci crystal), this band is the one with $b=p-1$, just before the highest band. Thus, as $\beta$ is increased, the value $\sigma_b=1$ of the Chern integer is ``transferred" from band $b$ to band $b+1$, starting from $b=(p+1)/2$ and ending at $b=p-1$. There are, therefore, $(p-3)/2$ transfers or topological phase transitions indicated by vertical bars in Fig. 4; these transitions are due to band degeneracies, see also Fig. 5.

\begin{figure}[tbp]
\includegraphics[width=8.2cm,trim = {1cm 1cm 0cm 0cm}]{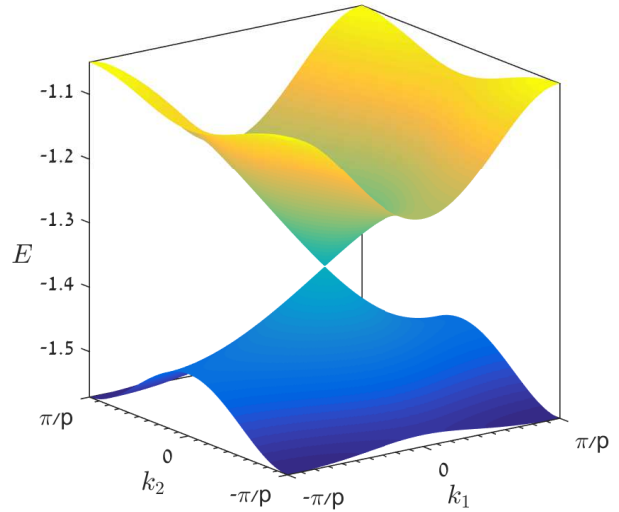}
\caption{(Color online) Plot of the second degeneracy at $\beta\approx 1.2144$ in Fig. 4. This is the degeneracy between bands 7 and 8 (counting from the lowest band in Fig. 4) at the point ${\mathbf k}=(0,0)$. In accordance with the Von-Neumann-Wigner theorem \cite{vnw}, three parameters ($k_1$, $k_2$, and $\beta$) of the Hermitian matrix $\hat{M}({\mathbf k})$ [defined by Eqs. (\ref{eep}) and (\ref{bc})] must be varied to get a degeneracy. This plot also clearly illustrates the periodicity of the band functions $E_b({\mathbf k})$ with period $2\pi /p$ in both $k_1$ and $k_2$, as mentioned after Eq. (\ref{BZ1}) (see also Appendix B).}
\end{figure}

Thus, for sufficiently large $\beta$, the total Chern integer in the gap above the separatrix band is $\sigma =1$. As indicated already by the semiclassical approximation (\ref{Ep}), this is the largest gap in the spectrum, see Fig. 4. Figure 6 shows the ``Hofstadter butterfly" (plot of the spectra at arbitrary rational $\nu =q/p$) for increasing values of $\beta$. We see that already for $\beta \geq 10$ (Figs. 6(c) and 6(d)) the maximal gap for $\nu\sim 1/p$ is actually almost independent on $\nu$, unlike cases of smaller $\beta$ (Figs. 6(a) and 6(b)). Clearly, for all $\nu <1/2$, this gap is associated with a total Chern integer $\sigma =1$. Thus, because of the general sum rule $\sum_{b=1}^p\sigma_b =1$ \cite{dz}, the total Chern integer of all the bands above this gap is $\sigma =0$. The number of these bands is approximately $q$ and they correspond to the highest band in the case of $q=1$ ($\nu =1/p$). For $p\gg 1$, one can show \cite{ad} that the width of the latter band (approximately equal to the total width of the $\sim q$ bands corresponding to it) is entirely due to the discontinuity of the Fibonacci potential ($\beta =\infty$) and that this band extends from $E=E_-=4/3$ to $E=E_+=2(\sqrt{5}-1)\approx 2.4721$, i.e., its width is $E_+-E_-\approx 1.1388$. The average energy $(E_++E_-)/2\approx 1.9027$ of this band should be compared with the semiclassical approximation (\ref{Ep}).

\begin{figure}[tbp]
\includegraphics[width=8cm,trim = {1.5cm 1cm 1cm 0cm}]{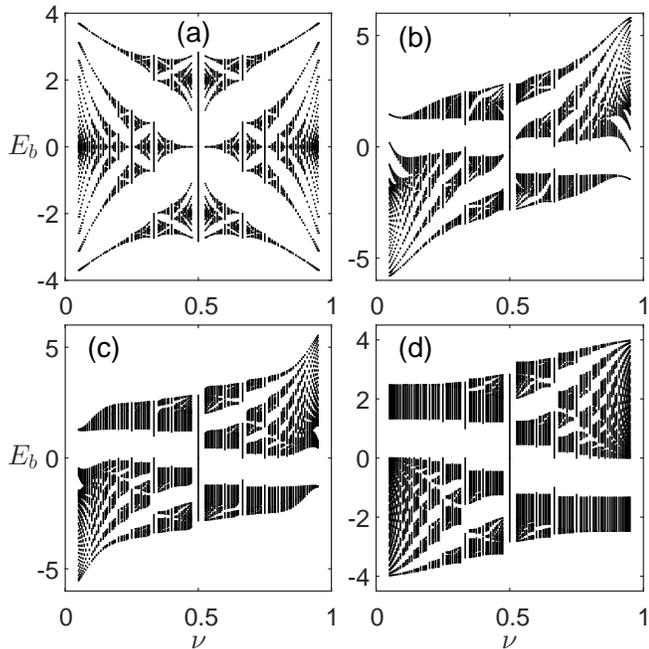}
\caption{(Color online) ``Hofstadter butterflies", i.e., energy bands as functions of rational values of $\nu$, for $\lambda = \lambda_\beta$ and (a) $\beta = 0.001$, (b) $\beta = 3$, (c) $\beta = 10$, and (d) $\beta = 1000$. In all cases, $\nu = q/p$ with $p = 1,\dots,20$ and, for each $p$, $q$ assumes all integer values that are relatively prime to $p$ and less than $p$. In case (a), close to the Harper limit of $\beta=0$, there is no dominant large gap in the semiclassical regime of $\nu\ll 1$. As $\beta$ is increased, a dominant large gap starts to form below the highest cluster of bands. In case (d), close to the Fibonacci limit of $\beta =\infty$, the width of this gap is relatively constant over a large interval of $\nu$, much beyond the semiclassical regime.}
\end{figure}

\begin{center}
\textbf{VI. SUMMARY AND CONCLUSIONS}
\end{center}

The Harper and generalized Fibonacci quasiperiodic systems (with irrational modulation frequency $\nu$) are known to exhibit energy spectra and eigenstates of basically different nature \cite{hm,ar,dh,aa,djt,hthk,tb0,tb1,tb2,tb3,tb4,tb5,tb6,tb7,tb8,tb9,tb10}. In this paper, we have studied the topological properties of these systems in their crystal (periodic) versions, i.e., for rational frequency $\nu =q/p$. By introducing an interpolating Hamiltonian (\ref{Heff}) with potential (\ref{Ub}) depending on a parameter $\beta$, the Harper crystal ($\beta =0$) can be transformed into a Fibonacci crystal ($\beta =\infty$). The basic topological differences between the two systems can be clearly seen in the semiclassical regime of $\nu\ll 1$, when the band energy spectrum is approximated by energy levels associated with classical orbits. The classical phase spaces of the two systems are significantly different, compare Figs. 3(a) and 3(c). For $\nu =1/p$, with $p$ odd and sufficiently large, the only band with nonzero Chern integer, $\sigma_b=1$, is the one whose energy is closest to the energy (\ref{Es}) of the classical separatrix orbit. The latter orbit is indeed topologically different from the contractible closed orbits associated with all other bands, with $\sigma_b=0$. As $\beta$ is increased from $\beta =0$, the separatrix energy (\ref{Es}) varies and, as a result, the band with $\sigma_b=1$ is shifted from the center of the spectrum [$b=(p+1)/2$], for $\beta=0$, to the band just below the highest one ($b=p-1$) for $\beta=\infty$; see Fig. 4. There occur, therefore, a relatively large number [$(p-3)/2$] of topological phase transitions as $\beta$ is varied from $\beta=0$ to $\beta =\infty$. 

It is interesting to compare the pure case \cite{dh} of $\nu =1/p$ with that of irrational $\nu$ for which no topological phase transition can occur (see Introduction). One would like to understand how the topological phase transitions for $\nu =1/p$ gradually disappear when approaching an irrational value of $\nu$, close to $1/p$, by its rational approximants. To this end, we use methods in Refs. \cite{mw,qc4,qc5} which we illustrate by the example of $\nu =1/[5+(\sqrt{5}-1)/2]$ (close to $\nu =1/5$), whose first approximants are $1/5$ and $2/11$. For $\nu =1/5$, one has 5 bands and $(p-3)/2=1$ topological phase transition occurring at $\beta\approx 1.2$, see Fig. 7(a). For $\nu =2/11$ and general $\beta$ (see Fig. 7(b)), the $11$ (sub)bands are grouped into $5$ clusters splitting from the $5$ bands for $\nu =1/5$. The total Chern number of a cluster is the same as that of the corresponding band for $\nu =1/5$. Each of the four clusters splitting from a ($\nu =1/5$, $\sigma_b=0$) band consists of two subbands with $\sigma_b=\pm 1$. The cluster splitting from the ($\nu =1/5$, $\sigma_b=1$) band consists of three subbands with $\sigma_b=\pm 1,1$. For $\beta\ll 1$, this is the third cluster, at the center of the spectrum. However, as $\beta$ is increased the upper subband of this cluster, indicated by an arrow in Fig. 7(b) and having $\sigma_b=1$, leaves the cluster and joins the fourth cluster. The total Chern number of the latter cluster thus changes from $0$ to $1$ while that of the third cluster changes from $1$ to $0$. This is in accordance with Fig. 7(a) but now these changes occur \emph{without any} topological phase transitions due to degeneracies. In fact, the only degeneracies in Fig. 7(b) occur \emph{within} clusters, so that the cluster Chern number does not change. But even the changes in the Chern numbers of the subbands in the latter clusters can be explained, at the next approximant level ($\nu =3/17$), by the ``motion" of one sub-subband from one subcluster to a neighboring one without the occurrence of any degeneracy.

\begin{figure}[tbp]
\includegraphics[width=8.2cm,trim = {1cm 1cm 0cm 0cm}]{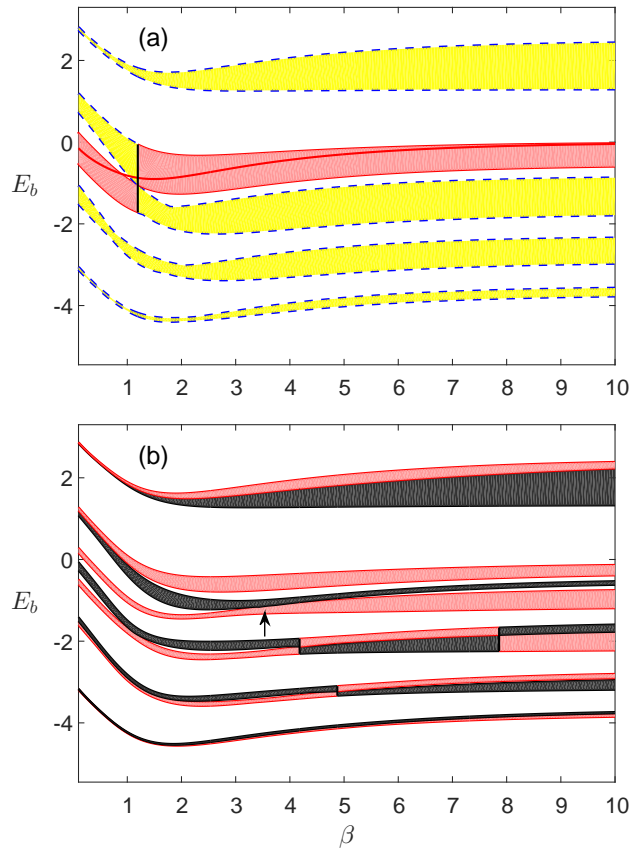}
\caption{(Color online) (a) Similar to Fig. 4 but for $\nu=1/5$, with only one band degeneracy at $\beta\approx 1.2$. (b) Case of $\nu =2/11$, featuring $11$ subbands that are grouped into $5$ clusters splitting from corresponding bands in (a). The red (gray) subbands have $\sigma_b=1$ while the dark gray subbands have $\sigma_b=-1$. See more details in text.}
\end{figure}

The problem of Bloch electrons in a magnetic field can be considered on an infinitely long strip of some width \cite{yh}. For a sufficiently wide strip and weak periodic potential at fixed Landau level (see Appendix A), this problem corresponds essentially to the truncation of Eq. (\ref{tbc}) to $|n|\leq N$, i.e., $\psi_n=0$ for $|n|>N$. Then, for arbitrary $\nu$, there emerge edge states with in-gap energy depending on a phase such as $\tau$ in Eq. (\ref{tbc}) or $k_1$ in Eq. (\ref{eep}) [without Eq. (\ref{bc}), since $k_2$ is not a good quantum number now]. The winding number of the edge-state energy in the gap as $k_1$ is varied turns out to be equal to $-\sigma$, where $\sigma$ is the Chern integer of the gap for the infinite system ($N=\infty$) \cite{yh}. In this way, the edge states for a strip feature the topological properties of the infinite system.      

These properties have been experimentally observed and used in realizations of 1D quasiperiodic systems (irrational $\nu$) on a strip by ``photonic quasicrystals" (PQs) \cite{kz,kz2,kz3}, i.e., finite quasiperiodic lattices of coupled single-mode optical waveguides. Localized edge states were observed for both Harper models and ``off-diagonal" (OD) quasiperiodic systems, i.e., systems (\ref{tbc}) without the on-site potential but with hopping constants modulated by the potential (\ref{Ub}). The transfer of an edge state from one side of the strip to the other by varying $\tau$ or $k_1$ (with the edge-state energy traversing a gap) was experimentally achieved by adiabatic pumping of light in PQ realizations of OD Harper and Fibonacci models \cite{kz,kz3}. The topological non-equivalence (equivalence) of two OD systems with different (equal) irrational values of $\nu$ was experimentally demonstrated by observing the presence (absence) of topological phase transitions in PQ realizations of the systems \cite{kz2}.

We plan to study in future works the topological properties, including edge states, of systems more general than those considered in this paper. In particular, OD systems and systems (\ref{Heff}) whose classical counterparts exhibit open orbits such as those in Fig. 2(a) for $\lambda\neq\lambda_{\beta}$. The study of the latter systems requires formalisms and methods much more complicated than those used in the present paper, see, e.g., Ref. \cite{qc6}. Results concerning the behavior of edge states as $\beta$ is varied should be useful in the experimental observation of topological phase transitions for rational $\nu$, such as those predicted in this paper.

\begin{center}
\textbf{ACKNOWLEDGMENTS}
\end{center}

The authors have benefited from useful discussions with Y.E.
Kraus, G. Murthy, and O. Zilberberg.

\begin{center}
\textbf{APPENDIX A}
\end{center}

For an electron with charge $-e$ and mass $M$ in a periodic potential $V(x,y)$ and a uniform magnetic field ${\bf B}$ in the $z$ direction, the Hamiltonian is $\hat{H}=\hat{\bf\Pi}^2/(2M)+V(\hat{x},\hat{y})$, where $\hat{\bf\Pi}= \hat{\bf p}+e{\bf B}\times\hat{\bf r}/(2c)$ is the kinetic momentum in the symmetric gauge. The operator $\hat{\bf\Pi}_{\rm c}= \hat{\bf p}-e{\bf B}\times\hat{\bf r}/(2c)$ gives the cyclotron-orbit center $(\hat{X},\hat{Y})=c/(eB)(-\hat{\Pi}_{{\rm c},y},\hat{\Pi}_{{\rm c},x})$  \cite{dz,jl}. We also define $(\hat{u},\hat{v})=c/(eB)(\hat{\Pi}_{y},\hat{\Pi}_{x})$ and assume, for simplicity, a $2\pi\times 2\pi$ unit cell of periodicity for the periodic potential $V$. The number $\varphi$ of flux quanta $\phi_0=hc/e$ per unit cell is given by $4\pi^2B=\varphi hc/e$, so that $\hbar c/(eB)=2\pi\nu$, where $\nu =1/\varphi$. We then get the commutation relations \cite{dz,jl}
\begin{equation}\label{cr1}
[\hat{u},\hat{v}]=2\pi i\nu,\ \ \ [\hat{X},\hat{Y}]=2\pi i\nu,
\end{equation}
\begin{equation}\label{cr2} 
[\hat{u},\hat{X}]=[\hat{u},\hat{Y}]=[\hat{v},\hat{X}]=[\hat{v},\hat{Y}]=0.
\end{equation}
Thus, $(\hat{u},\hat{v})$ and $(\hat{X},\hat{Y})$ are two independent conjugate pairs of variables. After expressing $\hat{\bf r}$ and $\hat{\bf p}$ in terms of these pairs, the Hamiltonian above reads as follows \cite{dz}
\begin{equation}
\hat{H}=\frac{M\omega^2}{2}(\hat{u}^2+\hat{v}^2) + V(\hat{X}+\hat{u},\hat{Y}-\hat{v}).
\end{equation}  
This describes a harmonic oscillator in $(\hat{u},\hat{v})$, with cyclotron frequency $\omega=eB/(Mc)$, perturbed by the potential $V$. If this perturbation is sufficiently weak relative to the spacing $\hbar\omega$ between oscillator (Landau) levels, matrix elements of $V$ between different Landau levels may be neglected. One can then write the $(u,X)$ representation of the eigenstate corresponding to the perturbed Landau level $l$ (for integer $l\geq 0$), with energy $E_l=(l+1/2)\hbar\omega$, as $F_l(u)G(X)$; here $F_l(u)$ is a normalized oscillator function and $G(X)$ satisfies the eigenvalue equation
\begin{equation}\label{Heee}
\hat{H}_{\mathrm{eff}}(\hat{X},\hat{Y})G(X)=EG(X),
\end{equation}
where
\begin{eqnarray}
&& \hat{H}_{\mathrm{eff}}(\hat{X},\hat{Y}) =\int_{-\infty}^{\infty}duF_l^{*}(u)V(\hat{X}+u,\hat{Y}-\hat{v})F_l(u)  \nonumber \\ && =\int_{-\infty}^{\infty}duF_l^{*}(u)V\left(\hat{X}+u,\hat{Y}+2\pi i\nu\frac{d}{du}\right)F_l(u), \label{Hee}
\end{eqnarray}
after using $\hat{v}=-2\pi i\nu d/du$ from Eq. (\ref{cr1}). The energy $E$ in Eq. (\ref{Heee}) is measured relative to $E_l$.

We now focus on the case of $V(x,y)=2\lambda W(x)+2\chi\cos(y)$, where $W(x)$ is $2\pi$-periodic and $(\lambda,\chi)$ are some real constants. By expanding $W(x)$ in a Fourier series, $W(x)=\sum_{j=-\infty}^{\infty}W_j\exp (ijx)$, one gets from Eq. (\ref{Hee}):
\begin{equation}\label{Hees}
\hat{H}_{\mathrm{eff}}(\hat{X},\hat{Y})=2\lambda U(\hat{X})+\kappa \left[e^{i\hat{Y}+i\alpha}+e^{-i\hat{Y}-i\alpha}\right], 
\end{equation} 
where
\begin{equation}\label{UX}
U(\hat{X})=\sum_{j=-\infty}^{\infty}W_j\int_{-\infty}^{\infty}du|F_l(u)|^2e^{iju}\ e^{ij\hat{X}},
\end{equation}
\begin{equation}\label{ka}
\kappa\exp (-i\alpha)=\chi \int_{-\infty}^{\infty}duF_l^{*}(u)F_l(u-2\pi\nu).
\end{equation}
Using $\hat{Y}=-2\pi i\nu d/dX$ from Eq. (\ref{cr1}), the constant phase $\alpha$ in Eq. (\ref{Hees}) can be removed by writing $G(X)=\exp[-i\alpha X/(2\pi\nu)]\bar{G}(X)$. Then, choosing $\chi$ in Eq. (\ref{ka}) so that $\kappa =1$, we see that Eq. (\ref{Heee}) is satisfied with $G(X)$ replaced by $\bar{G}(X)$ and with $\hat{H}_{\mathrm{eff}}(\hat{X},\hat{Y})$ given by Eq. (\ref{Heff}).

The latter equation for $\bar{G}(X)$ reads as follows:
\begin{equation}\label{Heeee}
\bar{G}(X+2\pi\nu )+\bar{G}(X-2\pi\nu )+2\lambda U(X)\bar{G}(X)=E\bar{G}(X).
\end{equation}  
Writing $X=X_0+2\pi n\nu$ for all integers $n$ and defining $\psi_n=\bar{G}(X_0+2\pi n\nu )$, Eq. (\ref{Heeee}) reduces to Eq. (\ref{tbc}).
  
\begin{center}
\textbf{APPENDIX B}
\end{center}

We show here that Eq. (\ref{Bs}) give Bloch eigenstates of $\hat{H}_{\mathrm{eff}}(\hat{X},\hat{Y})$ and derive some properties of these states. We also derive the Diophantine equation (\ref{de}) for the topological integers. We first note that the Hamiltonian (\ref{Heff}) commutes with translations by $2\pi$ in $(\hat{X},\hat{Y})$:
\begin{equation}\label{dxy}
\hat{D}_{X,2\pi}=\exp(i\hat{Y}/\nu ),\ \ \hat{D}_{Y,2\pi}=\exp(-i\hat{X}/\nu ),
\end{equation}
where we used $\hat{Y}=-2\pi i\nu d/dX$ and $\hat{X}=2\pi i\nu d/dY$ (from $[\hat{X},\hat{Y}]=2\pi i\nu$). In general, the translations (\ref{dxy}) do not commute:
\begin{equation}\label{dxdy}
\hat{D}_{X,2\pi}\hat{D}_{Y,2\pi}=\exp(-2\pi i/\nu )\hat{D}_{Y,2\pi}\hat{D}_{X,2\pi} .
\end{equation} 
However, for rational $\nu =q/p$, some powers of the translations (\ref{dxy}) will commute. For example, using Eq. (\ref{dxdy}), one can see that $\hat{D}_{X,2\pi}^q=\hat{D}_{X,2\pi q}$ is the smallest translation of $\hat{X}$ by a multiple ($q$) of $2\pi$ that commutes with $\hat{D}_{Y,2\pi}$. It is then easy to check that the states (\ref{Bs}) are eigenstates of $\hat{D}_{X,2\pi q}$, $\hat{D}_{Y,2\pi}$, and $\hat{H}_{\mathrm{eff}}(\hat{X},\hat{Y})$ with respective eigenvalues $\exp (ipk_2)$, $\exp (-ik_1/\nu )$, and $E_b({\mathbf k})$.   

It follows from Sec. IIIA that for an isolated band $b$ $E_b({\mathbf k})$ is periodic with two zones of periodicity, Eqs. (\ref{BZ}) and (\ref{BZ1}). This is possible only if $E_b({\mathbf k})$ is periodic in both $k_1$ and $k_2$ with period $2\pi /p$, defining a periodicity zone $q$ times smaller than the BZ (\ref{BZ}). This means that the eigenstates (\ref{Bs}) are $q$-fold degenerate. In fact, since $\hat{D}_{X,2\pi}$ in Eq. (\ref{dxy}) commutes with $\hat{H}_{\mathrm{eff}}(\hat{X},\hat{Y})$, the $q$ states $\hat{D}_{X,2\pi}^s\Psi_{b,{\mathbf k}}(X)$, $s=0,...,q-1$, are all degenerate in energy; using Eq. (\ref{dxdy}), one finds that these states are eigenstates in band $b$ associated with the quasimomenta ${\mathbf k}_s=(k_1-2\pi s\ {\rm mod}(2\pi\nu),k_2)$. 

We now derive the Diophantine equation (\ref{de}). As mentioned at the end of Sec. III, the total phase change of $\Psi_{b,{\mathbf k}}(X)$ when going around the boundary of the BZ (\ref{BZ}) counterclockwise must be an integer multiple of $2\pi$ and we denote this integer by $-\sigma_b$. From Eqs. (\ref{pc1}) and (\ref{pc2}), it follows that when $k_2$ is varied from $k_2$ to $k_2+2\pi /p$ (on the vertical axis of the BZ), the total change in the phase of  $\Psi_{b,k_1+2\pi\nu,k_2}(X)$ relative to that of $\Psi_{b,{\mathbf k}}(X)$ is $f_b(k_1,k_2+2\pi /p)-f_b({\mathbf k})$; when $k_1$ is varied from $k_1+2\pi\nu$ to $k_1$ (on the horizontal axis of the BZ), the total change in the phase of $\Psi_{b,k_1,k_2+2\pi /p}(X)$ relative to that of $\Psi_{b,{\mathbf k}}(X)$ is $g_b({\mathbf k})-g_b(k_1+2\pi\nu,k_2)$. Thus, one must have
\begin{eqnarray}\label{ci1}
f_b(k_1,k_2+2\pi /p) & - & f_b({\mathbf k}) \\ \notag & + & g_b({\mathbf k}) -g_b(k_1+2\pi\nu,k_2) =-2\pi \sigma_b . 
\end{eqnarray}
Similarly, the total phase change of ${\mathbf V}_b({\mathbf k})$ when going around the boundary of zone (\ref{BZ1}) counterclockwise must be $2\pi\mu_b$, where $\mu_b$ is a second Chern integer determined from Eqs. (\ref{vpc1}) and (\ref{vpc2}):           
\begin{eqnarray}\label{ci2}
w_b(k_1,k_2+2\pi /p) & - & w_b({\mathbf k}) \\ \nonumber & + & g_b({\mathbf k})-g_b(k_1+2\pi ,k_2) =2\pi\mu_b . 
\end{eqnarray}

Now, let us iterate Eq. (\ref{pc1}) $p$ times. This gives
\begin{equation}\label{pcp}
\Psi_{b,k_1+2\pi q,k_2}(X) = \exp [i\bar{f}_b({\mathbf k})]\Psi_{b,{\mathbf k}}(X),
\end{equation}
where 
\begin{equation}\label{bfb}
\bar{f}_b({\mathbf k})=\sum_{r=0}^{p-1}f_b(k_1+2\pi r\nu ,k_2).
\end{equation} 
On the other hand, using Eq. (\ref{Bs}) with Eq. (\ref{kq}) and the $q$th iteration of Eq. (\ref{vpc1}), we get
\begin{equation}\label{pcpn}
\Psi_{b,k_1+2\pi q,k_2}(X) = \exp [i\bar{w}_b({\mathbf k})-ipk_2]\Psi_{b,{\mathbf k}}(X),
\end{equation}
where
\begin{equation}\label{bwb}
\bar{w}_b({\mathbf k})=\sum_{r=0}^{q-1}w_b(k_1+2\pi r,k_2)
\end{equation} 
and $w_b({\mathbf k})$ are the phases in Eq. (\ref{vpc1}). Then, by comparing Eq. (\ref{pcp}) with Eq. (\ref{pcpn}), it follows that
\begin{equation}\label{epcp}
\bar{w}_b({\mathbf k})-pk_2=\bar{f}_b({\mathbf k})+2\pi z,
\end{equation}
where $z$ is some integer. Next, let us add $p$ equations (\ref{ci1}) for $k_1,k_1+2\pi\nu,\dots,k_1+2\pi\nu (p-1)$. Using Eq. (\ref{bfb}), we find that
\begin{eqnarray}\label{cip}
\bar{f}_b(k_1,k_2+2\pi /p) & - & \bar{f}_b({\mathbf k}) \\ \notag & + & g_b({\mathbf k}) -g_b(k_1+2\pi q,k_2) =-2\pi p\sigma_b . 
\end{eqnarray}
Similarly, by adding $q$ equations (\ref{ci2}) for $k_1,k_1+2\pi ,\dots,k_1+2\pi (q-1)$  and using Eq. (\ref{bwb}), we get 
\begin{eqnarray}\label{ciq}
\bar{w}_b(k_1,k_2+2\pi /p) & - & \bar{w}_b({\mathbf k}) \\ \notag & + & g_b({\mathbf k}) -g_b(k_1+2\pi q,k_2) =2\pi q\mu_b . 
\end{eqnarray}
Because of Eq. (\ref{epcp}), Eq. (\ref{cip}) can be rewritten as 
\begin{eqnarray}\label{cipq}
& &\bar{w}_b(k_1,k_2+2\pi /p) - \bar{w}_b({\mathbf k})-2\pi \\ \notag & + & g_b({\mathbf k}) -g_b(k_1+2\pi q,k_2) =-2\pi p\sigma_b . 
\end{eqnarray}       
Finally, by subtracting Eq. (\ref{cipq}) from Eq. (\ref{ciq}) and dividing by $2\pi$, we obtain Eq. (\ref{de}).

\begin{center}
\textbf{APPENDIX C}
\end{center}

The classical Hamilton equations for the Hamiltonian (\ref{Heff}) are:
\begin{equation} \label{hes}
\dot{X}=\frac{\partial H_{\mathrm{eff}}}{\partial Y},\ \ \
\dot{Y}=-\frac{\partial H_{\mathrm{eff}}}{\partial X}.  
\end{equation}
The fixed points of the motion are determined from $\dot{X}=\dot{Y}=0$. Using Eqs. (\ref{Heff}) and (\ref{hes}), we get from $\partial H_{\mathrm{eff}}/\partial X=0$ that $\sin (Y)=0$ or $Y=0,\pi$. Similarly, from Eqs. (\ref{Heff}) and (\ref{Ub}), we find   
\begin{equation}\label{xdot0}
\frac{\partial H_{\mathrm{eff}}}{\partial Y}=\frac{2\lambda\sin(X)}{\tanh(\beta)\cosh^2\{\beta[\cos(X)-\gamma_{\nu}]\}},
\end{equation}
so that $\partial H_{\mathrm{eff}}/\partial Y=0$ implies $\sin (X)=0$ or $X=0,\pi$. There are therefore four fixed points, $(X,Y)=(0,0),(\pi,0),(0,\pi),(\pi,\pi)$, within the $2\pi\times 2\pi$ unit cell of periodicity of the Hamiltonian (\ref{Heff}). Let us determine the linear stability of these points, $\mathbf{R}\equiv (X,Y)$ (a column vector), under small perturbations $\delta\mathbf{R}$ around them. By linearizing Eqs. (\ref{hes}) around $\mathbf{R}$, we get 
\begin{equation}\label{lhe}
\dot{\delta\mathbf{R}}=DH_{\mathrm{eff}} \delta\mathbf{R},
\end{equation}
where $DH_{\mathrm{eff}}$ is the matrix 
\begin{equation}\label{dhe}
DH_{\mathrm{eff}}= 
\begin{pmatrix}
\frac{\partial ^{2}H_{\mathrm{eff}}}{\partial X\partial Y} & \frac{
\partial ^{2}H_{\mathrm{eff}}}{\partial Y^{2}} \\ 
&  \\ 
-\frac{\partial ^{2}H_{\mathrm{eff}}}{\partial X^{2}} & -\frac{\partial
^{2}H_{\mathrm{eff}}}{ \partial X\partial Y}
\end{pmatrix}
.
\end{equation} 
Assuming the time dependence $\delta\mathbf{R}(t)=\delta\mathbf{R}_0\exp(\xi t)$
in Eq. (\ref{lhe}), we get the eigenvalue equation 
\begin{equation}\label{ee}
DH_{\mathrm{eff}} \delta\mathbf{R}_0=\xi \delta\mathbf{R}_0.
\end{equation}
Since the matrix (\ref{dhe}) has vanishing trace, the two eigenvalues satisfy $\xi_1+\xi_2=0$. The fixed points are unstable only if these eigenvalues are real. A simple calculation of $\xi_{1,2}$ using Eq. (\ref{dhe}) with Eqs. (\ref{Heff}) and (\ref{Ub}) shows that this is the case only for the points $\mathbf{R}=(\pi,0),(0,\pi)$. From each of these points there emanates a separatrix orbit (see, e.g., Fig. 2(a)) whose energy is the energy $E=H_{\mathrm{eff}}(X,Y)$ of the point. However, if the two points have the same energy, there will be only one separatrix connecting both points, as in the case of Fig. 2(b). After calculating $E=H_{\mathrm{eff}}(X,Y)$ for the two points and requiring that $H_{\mathrm{eff}}(\pi,0)=H_{\mathrm{eff}}(0,\pi)$, we obtain the results (\ref{lc}) and (\ref{Es}).


\begin{thebibliography}{99}
\bibitem{tknn} D.J. Thouless, M. Kohmoto, M.P. Nightingale, and M. den Nijs,
Phys. Rev. Lett. \textbf{49}, 405 (1982).

\bibitem{ass} J.E. Avron, R. Seiler, and B. Simon, Phys. Rev. Lett. \textbf{51}, 51 (1983).

\bibitem{bs} B. Simon, Phys. Rev. Lett. \textbf{51}, 2167 (1983).

\bibitem{ahm} A.H. MacDonald, Phys. Rev. B \textbf{29}, 3057 (1984).

\bibitem{mk} M. Kohmoto, Ann. Phys. \textbf{160}, 343 (1985).

\bibitem{daz} I. Dana, Y. Avron, and J. Zak, J. Phys. C \textbf{18}, L679
(1985).

\bibitem{dz} I. Dana and J. Zak, Phys. Rev. B \textbf{32}, 3612 (1985), and
references therein; Phys. Lett. A \textbf{146}, 147 (1990).

\bibitem{kunz} H. Kunz, Phys. Rev. Lett. \textbf{57}, 1095 (1986).

\bibitem{id0} I. Dana, Phys. Lett. A \textbf{150}, 253 (1990).

\bibitem{hk} Y. Hatsugai and M. Kohmoto, Phys. Rev. B \textbf{42}, 8282
(1990).

\bibitem{yh} Y. Hatsugai, Phys. Rev. Lett. \textbf{71}, 3697 (1993); Phys. Rev. B \textbf{48}, 11851 (1993).

\bibitem{kz} Y.E. Kraus, Y. Lahini, Z. Ringel, M. Verbin, and O. Zilberberg,
Phys. Rev. Lett. \textbf{109}, 106402 (2012). 

\bibitem{kz1} Y.E. Kraus and O. Zilberberg, Phys. Rev. Lett. \textbf{109},
116404 (2012).

\bibitem{kz2} M. Verbin, O. Zilberberg, Y.E. Kraus, Y. Lahini, and Y.
Silberberg, Phys. Rev. Lett. \textbf{110}, 076403 (2013).

\bibitem{mbb} K.A. Madsen, E.J. Bergholtz, and P.W. Brouwer, Phys. Rev. B \textbf{88}, 125118 (2013).

\bibitem{id1} I. Dana, Phys. Rev. B \textbf{89}, 205111 (2014).

\bibitem{kz3} M. Verbin, O. Zilberberg, Y. Lahini, Y.E. Kraus, and Y. Silberberg,  Phys. Rev. B  \textbf{91}, 064201 (2015).

\bibitem{kz4} Y.E. Kraus and O. Zilberberg, Nature Physics \textbf{12}, 624 (2016). 

\bibitem{mw} M. Wilkinson, J. Phys. A \textbf{20}, 4337 (1987); \textbf{27}, 8123 (1994).

\bibitem{qc1} P. Leboeuf, J. Kurchan, M. Feingold, and D.P. Arovas, Phys.
Rev. Lett. \textbf{65}, 3076 (1990); Chaos \textbf{2}, 125 (1992).

\bibitem{qc2} F. Faure and P. Leboeuf, in \emph{Proceedings of the
Conference: From Classical to Quantum Chaos}, edited by G.F. Dell'Antonio,
S. Fantoni, and V.R. Manfredi (SIF, Bologna, 1993), Vol. 41.

\bibitem{qc3} I. Dana, Phys. Rev. E \textbf{52}, 466 (1995).

\bibitem{qc4} I. Dana, M. Feingold, and M. Wilkinson, Phys. Rev. Lett. 
\textbf{81}, 3124 (1998).

\bibitem{qc5} I. Dana, Y. Rutman, and M. Feingold, Phys. Rev. E \textbf{58},
5655 (1998).

\bibitem{qc6} F. Faure, J. Phys. A \textbf{33}, 531 (2000).

\bibitem{fti} T. Kitagawa, E. Berg, M. Rudner, and E. Demler, Phys. Rev. B {\bf 82}, 235114 (2010); N.H. Lindner, G. Refael, and V. Galitski, Nature Phys. {\bf 7}, 490 (2011); Z. Gu, H.A. Fertig, D.P. Arovas, and A. Auerbach, Phys. Rev. Lett. {\bf 107}, 216601 (2011); T. Kitagawa, T. Oka, A. Brataas, L. Fu, and E. Demler, Phys. Rev. B {\bf 84}, 235108 (2011); M.S. Rudner, N.H. Lindner, E. Berg, and M. Levin, Phys. Rev. X {\bf 3}, 031005 (2013).

\bibitem{lsz} M. Lababidi, I.I. Satija, and E. Zhao, Phys. Rev. Lett. {\bf 112}, 026805 (2014).

\bibitem{jg1} D.Y.H. Ho and J.B. Gong, Phys. Rev. Lett. {\bf 109}, 010601 (2012).

\bibitem{id2} I. Dana, Phys. Rev. E {\bf 96}, 022216 (2017).

\bibitem{hm} P.G. Harper, Proc. Phys. Soc. London Sect. A \textbf{68}, 874
(1955).

\bibitem{ar} A. Rauh, Phys. Status Solidi B \textbf{65}, K131 (1974); 
\textit{ibid.} \textbf{69}, K9 (1975).

\bibitem{dh} D.R. Hofstadter, Phys. Rev. B \textbf{14}, 2239 (1976), and
references therein.

\bibitem{aa} S. Aubry and G. Andr\'{e}, Ann. Isr. Phys. Soc. \textbf{3}, 133
(1980).

\bibitem{djt} D.J. Thouless, Phys. Rev. B \textbf{28}, 4272 (1983); Commun.
Math. Phys. \textbf{127}, 187 (1990).

\bibitem{hthk} J.H. Han, D.J. Thouless, H. Hiramoto, and M. Kohmoto, Phys.
Rev. B \textbf{50}, 11 365 (1994).

\bibitem{tb0} H. Hiramoto and M. Kohmoto, Int. J. Mod. Phys. B \textbf{06},
281 (1992), and references therein.

\bibitem{tb1} M. Kohmoto, L.P. Kadanoff, and C. Tang, Phys. Rev. Lett. 
\textbf{50}, 1870 (1983).

\bibitem{tb2} S. Ostlund, R. Pandit, D. Rand, H.J. Schellnhuber, and E.D.
Siggia, Phys. Rev. Lett. \textbf{50}, 1873 (1983).

\bibitem{tb3} J. Bellissard, B. Iochum, E, Scoppola, and D. Testard, Commun.
Math. Phys. \textbf{125}, 527 (1989).

\bibitem{tb4} Y. Liu and R. Riklund, Phys. Rev. B \textbf{35}, 6034 (1987).

\bibitem{tb5} Q. Niu and F. Nori, Phys. Rev. B \textbf{42}, 10329 (1990).

\bibitem{tb6} Y. Liu and W. Sritrakool, Phys. Rev. B \textbf{43}, 1110
(1991), and references therein.

\bibitem{tb7} G. Gumbs and M.K. Ali, Phys. Rev. Lett. \textbf{60}, 1081
(1988); J. Phys. A \textbf{22}, 951 (1989).

\bibitem{tb8} J.Q. You, J.R. Yan, T. Xie, X. Zeng, and
J.X. Zhong, J. Phys.: Condens. Matter \textbf{3}, 7255 (1991).

\bibitem{tb9} G.Y. Oh, C.S. Ryu, and M.H. Lee, Phys. Rev. B \textbf{47},
6122 (1993), and references therein.

\bibitem{tb10} D. Damanik and A. Gorodetski, Commun. Math. Phys. \textbf{305}, 221 (2011).

\bibitem{jz} J. Zak, in \emph{Solid State Physics}, edited by H. Ehrenreich, F. Seitz, and D. Turnbull (Academic, New York, 1972), Vol. 27, and references therein. 

\bibitem{ad} G. Amit and I. Dana, to be published.

\bibitem{note} As explained after Eq. (\ref{dxdy}) in Appendix B, minimal commuting phase-plane translations for $\nu =q/p$ are $\hat{D}_{X,2\pi q}$ and $\hat{D}_{Y,2\pi}$. These define a ``quantum" unit cell $q$ times larger than the classical $2\pi\times 2\pi$ unit cell of periodicity. Then, the case of $q>1$ is more complicated than that of $q=1$ since one has to consider tunneling effects between the $q$ classical cells within the quantum cell in order to establish a classical-quantum correspondence in the semiclassical regime \cite{qc2}. Also, for even $p$, there may occur band degeneracies for all $\beta$, so that we consider here only the case of odd $p$ for simplicity.

\bibitem{vnw} J. von Neumann and E.P. Wigner, Z. Physik \textbf{30}, 467 (1929).

\bibitem{jl} M.H. Johnson and B.A. Lippmann, Phys. Rev. \textbf{76}, 828 (1949).

\end{thebibliography}
\end{document}